\documentclass[conference]{IEEEtran}

\usepackage{cite}
\usepackage{amsmath, amssymb, amsfonts}
\usepackage{graphicx}
\usepackage{subfig}
\usepackage{textcomp}
\usepackage{xcolor}

\usepackage{algorithm}
\usepackage{algpseudocode}

\usepackage{balance}

\usepackage{amsthm}

\begin{document}
\title{QoS- and Physics-Aware Routing in Optical LEO Satellite Networks \\ via Deep Reinforcement Learning}

\author{Mohammad~Taghi~Dabiri,~Rula~Ammuri, 
	Mazen~Hasna,~{\it Senior Member,~IEEE},\\ 
	~Khalid~Qaraqe,~{\it Senior Member,~IEEE}
	\thanks{M.T. Dabiri is with the College of Science and Engineering, Hamad Bin Khalifa University, Doha, Qatar. email: (mdabiri@hbku.edu.qa).}
	
	\thanks{R. Ammuri is with Professionals for Smart Technology (PST), Amman, Jordan (email: rammuri@pst.jo).}
	
	\thanks{M. Hasna is with the Department of Electrical Engineering, Qatar University, Doha, Qatar (e-mail: hasna@qu.edu.qa).}
	
	\thanks{Khalid A. Qaraqe is with the College of Science and Engineering, Hamad Bin Khalifa University, Doha, Qatar, and also with the Department of Electrical Engineering, Texas A\&M University at Qatar, Doha, Qatar (e-mail: kqaraqe@hbku.edu.qa).}
	
	\thanks{This publication was made possible by NPRP14C-0909-210008 from the Qatar Research, Development and Innovation (QRDI) Fund (a member of Qatar Foundation), Texas A\&M University at Qatar, and Hamad Bin Khalifa University, which supported this publication. }
}

\maketitle
\begin{abstract}
	Optical inter-satellite links (ISLs) are becoming the principal communication backbone in modern large-scale LEO constellations, offering multi-Gb/s capacity and near speed-of-light latency. However, the extreme sensitivity of optical beams to relative satellite motion, pointing jitter, and rapidly evolving geometry makes routing fundamentally more challenging than in RF-based systems. In particular, intra-plane and inter-plane ISLs exhibit markedly different stability and feasible range profiles, producing a dynamic, partially constrained connectivity structure that must be respected by any physically consistent routing strategy. This paper presents a lightweight geometry- and QoS-aware routing framework for optical LEO networks that incorporates class-dependent feasibility constraints derived from a jitter-aware Gaussian-beam model. These analytically computed thresholds are embedded directly into the time-varying ISL graph and enforced via feasible-action masking in a deep reinforcement learning (DRL) agent. The proposed method leverages local geometric progress, feasible-neighbor structure, and congestion indicators to select next-hop relays without requiring global recomputation. Simulation results on a Starlink-like constellation show that the learned paths are physically consistent, exploit intra-plane stability, adapt to jitter-limited inter-plane connectivity, and maintain robust end-to-end latency under dynamic topology evolution.
\end{abstract}

\begin{IEEEkeywords}
	Low Earth orbit (LEO) satellite networks, optical inter-satellite links (ISLs), QoS-aware routing, deep reinforcement learning (DRL).
\end{IEEEkeywords}

\IEEEpeerreviewmaketitle


\section{Introduction}
The rapid deployment of large low-Earth-orbit (LEO) satellite constellations marks a major transformation in global broadband connectivity. Modern systems such as Starlink and OneWeb increasingly rely on dense mesh topologies formed by optical inter-satellite links (ISLs), enabling multi-gigabit space-to-space communication, globally consistent latency, and reduced reliance on terrestrial backbone infrastructure \cite{salim2024cybersecurity,wu2025enhancing}. Compared with traditional RF crosslinks, optical ISLs provide narrower beams, stronger interference isolation, and significantly higher spectral efficiency. These advantages make optical ISLs a natural candidate for supporting emerging low-latency applications such as distributed cloud services, immersive media, and sensing-integrated communications. However, the same properties that enable high throughput also introduce a number of stringent physical constraints that must be incorporated into both network design and routing protocols.

First, optical ISLs require tight beam alignment, and their performance is highly sensitive to platform vibration, terminal misalignment, and residual pointing jitter. These effects lead to non-negligible pointing loss and distance-dependent outage probabilities \cite{10553228,10681506,dabiri2019optimal}, causing substantial variations in link reliability across the constellation. Second, the LEO environment inherently exhibits rapid geometric evolution: satellites travel at high orbital velocity, which causes continuous changes in inter-satellite distances, relative angular rates, and mutual visibility. As a result, the ISL connectivity graph evolves quickly over time, and end-to-end delay depends not only on geographic separation but also on the instantaneous orbital configuration. Third, intra-plane and inter-plane ISLs behave differently in practice. Intra-plane links benefit from nearly constant geometry and stable tracking, whereas inter-plane links experience higher relative motion, increased jitter, and more frequent link disruptions. This heterogeneous stability profile produces a time-varying feasible link set in which many candidate ISLs (especially cross-plane ones) may exceed their outage-limited operational range. Fourth, traffic patterns and queue build-up at certain satellites introduce local congestion, which can invalidate otherwise geometrically feasible routes and further reduce the set of acceptable next-hop choices.

The research community has investigated many of these challenges individually. Prior efforts include deterministic topology-aware routing \cite{cao2022dynamic}, grid and region-based routing \cite{yang2020low,li2023leo}, on-demand and distributed approaches \cite{stock2022distributed,bhattacharjee2024demand}, and load-aware and multi-plane routing mechanisms tailored to large-scale networks \cite{wang2023orbit,wang2024mldr,zhang2024link}. Reinforcement learning (RL)-based routing methods have also gained attention due to their ability to adapt to fast-moving topologies and implicitly learn complex decision structures \cite{dong2023drl,chen2025deep}. Meanwhile, laser-based ISL design and topology analysis have been explored in works such as \cite{yang2024analysis,10437539}. However, despite these advances, most existing schemes rely on simplified assumptions about ISL reachability, typically modeling ISLs as ideal, fixed-range links without directly incorporating the effects of pointing jitter, outage probability, or the distinct behaviors of intra-plane versus inter-plane connectivity. Consequently, prior routing strategies may select geometrically short paths that are physically infeasible or unstable when deployed on real optical terminals.

Three characteristics of optical LEO networks highlight the need for routing algorithms that explicitly account for physical feasibility. (i) ISL reliability is determined jointly by geometry, beam divergence, and terminal tracking accuracy rather than simple distance thresholds. (ii) As satellites move, the feasible ISL set can change significantly between snapshots, producing a dynamic and sometimes sparse connectivity structure. (iii) Latency-optimal routes often require combining long but stable intra-plane hops with strategically chosen inter-plane transitions that re-align the route with the source–destination great-circle arc. These factors collectively demand routing approaches that are both lightweight and physically informed, capable of exploiting geometric regularities while reacting to rapid changes in link feasibility and congestion.

\subsection{Contributions}
This work develops a physically grounded and computationally efficient routing framework for optical LEO constellations that captures class dependent ISL feasibility, fast geometric evolution, and dynamic traffic conditions. At the physical layer, we integrate the impact of pointing jitter, beam divergence, and class specific optical stability through precomputed maximum feasible ISL ranges. These thresholds distinguish between intra plane and inter plane links and are embedded directly in the snapshot graph, which enables fast feasibility checks without on the fly SNR evaluation and ensures that all routing decisions remain consistent with the underlying optical constraints.

On top of this feasibility model, we design a lightweight deep reinforcement learning (DRL) routing agent that uses local geometric progress, the structure of feasible neighbors, and simple congestion indicators to select next hop relays. Feasible action masking enforces all optical and traffic constraints during decision making, which reduces the exploration burden and improves routing stability in highly dynamic constellation snapshots. Using a Starlink like orbital configuration, we evaluate the proposed method and show that the learned routes are physically consistent, naturally exploit the higher stability of intra plane ISLs, adapt to the tighter ranges of jitter limited inter plane links, and maintain robust end to end latency under topology evolution. Overall, the paper presents a QoS and geometry aware routing framework that operates efficiently in dynamic optical LEO networks while respecting key physical constraints of modern spaceborne laser communication systems.

\section{System Model}
We consider a large-scale optical LEO satellite constellation equipped with inter-satellite laser terminals and a set of geographically distributed ground gateways. This section describes the geometric layout, line-of-sight conditions, optical link feasibility model, and delay structure that together form the physical foundation for the routing framework used in later sections.

\subsection{Constellation Geometry}
The satellites are deployed over multiple circular orbital planes at altitude $h$ above Earth with radius $R_e$. Let $r=R_e+h$ denote the orbital radius. Each orbital plane has inclination $\theta_{\mathrm{inc}}$ and right ascension of ascending node (RAAN) separated by a constant offset. The position of satellite $k$ in plane $p$ at time $t$ is represented as
\begin{align}
	\boldsymbol{r}_{p,k}(t)
	=
	r\,\mathbf{R}_z(\Omega_p)\,
	\mathbf{R}_x(\theta_{\mathrm{inc}})
	\begin{bmatrix}
		\cos(\omega t + \phi_{p,k})\\
		\sin(\omega t + \phi_{p,k})\\
		0
	\end{bmatrix},
\end{align}
where $\omega=\sqrt{\mu/r^3}$ is the orbital angular velocity, $\mu$ is Earth’s gravitational constant, and $\phi_{p,k}$ is the in-plane phase offset. Because satellites move continuously along their orbits, inter-satellite distances $\ell_{ij}(t)=\|\boldsymbol{r}_i(t)-\boldsymbol{r}_j(t)\|$ and visibility conditions evolve on the order of seconds.

\subsection{Gateway Visibility}
Each gateway $g$ at geodetic coordinates $(\varphi_g,\lambda_g)$ is mapped to
\begin{align}
	\boldsymbol{r}_g = R_e
	\begin{bmatrix}
		\cos\varphi_g\cos\lambda_g\\
		\cos\varphi_g\sin\lambda_g\\
		\sin\varphi_g
	\end{bmatrix}.
\end{align}
A satellite $i$ is allowed to serve gateway $g$ only if the elevation angle satisfies
\begin{align}
	\varepsilon_{i,g}(t) \ge \varepsilon_{\min},
\end{align}
which ensures clear line-of-sight connectivity and avoids low-angle fading or atmospheric blockage.

\subsection{Optical ISL Model}
Each inter-satellite link operates through a narrow-beam optical terminal with Gaussian intensity distribution \cite{andrews2023laser}. Let $\ell_{ij}(t)$ denote the separation between satellites $i$ and $j$. The received power under perfect alignment is modeled as
\begin{align}
	P_{r,0}(t)
	=
	P_t\,\eta_t\eta_r\left(\frac{a_r}{w(\ell_{ij})}\right)^2,
\end{align}
where $P_t$ is the transmit power, $(\eta_t,\eta_r)$ are optical efficiencies, $a_r$ is the receiver aperture radius, and $w(\ell)$ is the beam radius at distance $\ell$. A standard far-field approximation yields \cite{andrews2023laser}
\begin{align}
	w(\ell)\approx \theta_{\text{div}}\,\ell,
\end{align}
where $\theta_{\text{div}}$ is the beam divergence angle.

\subsubsection{Pointing Error Effects}
Optical ISLs in LEO are highly sensitive to mechanical vibration, platform jitter, and residual tracking error. We adopt a widely used model where the instantaneous pointing error is Gaussian with variance $\sigma_\theta^2$, which differs between intra-plane and inter-plane links due to their relative dynamics:
\begin{align}
	\sigma_{\theta,\text{inter}} > \sigma_{\theta,\text{intra}}.
\end{align}
The resulting pointing loss is \cite{dabiri2019optimal}
\begin{align}
	L_{\mathrm{pt}}(\theta)=\exp\Big(-2(\ell\theta/w(\ell))^2\Big),
\end{align}
and the instantaneous SNR on link $e$ becomes
\begin{align}
	\gamma_e(t)=\frac{P_{r,0}(t)L_{\mathrm{pt}}(\theta)}{N_0 B}.
\end{align}
A link is considered reliable if $\gamma_e(t)\ge\gamma_\mathrm{th}$ with outage probability below a threshold $P_{\mathrm{out,th}}$.

\subsubsection{Class-Dependent Feasible Range}
To avoid evaluating outage probability at runtime, we compute \emph{offline} the maximum feasible distance of each ISL class:
\begin{align}
	\ell_e(t)\le 
	\begin{cases}
		L_{\max}^{\text{intra}}, & \text{intra-plane links},\\[2pt]
		L_{\max}^{\text{inter}}, & \text{inter-plane links}.
	\end{cases}
	\label{eq:feasible_range_system}
\end{align}
These values incorporate optical power, aperture geometry, divergence, jitter levels, and SNR requirements. The routing layer only checks \eqref{eq:feasible_range_system}, making the feasibility test fast and independent of physical-layer parameters.

\subsection{Propagation Delay Model}
The propagation delay of ISL $e$ at time $t$ is given by
\begin{align}
	\tau_e(t)=\frac{\ell_e(t)}{c},
\end{align}
with $c$ the speed of light. The end-to-end propagation delay of a route $\mathcal{P}$ is thus
\begin{align}
	T_{\mathrm{prop}}(t)=\sum_{e\in\mathcal{P}}\frac{\ell_e(t)}{c}.
\end{align}
A constant forwarding overhead $\beta$ captures digital processing, buffering, and optical acquisition overhead. Hence, each hop incurs a total cost
\begin{align}
	w_e(t)=\frac{\ell_e(t)}{c}+\beta.
	\label{eq:link_cost}
\end{align}

\subsection{Feasible Link Construction}
Combining geometric visibility, optical feasibility, and class-dependent jitter behavior, the feasible link set at time $t$ is defined as
\begin{align}
	&\mathcal{E}_{\mathrm{feas}}(t)
	= 
	\Big\{
	e=(i\!\to\!j)\in\mathcal{E}(t):
	\nonumber \\&~~~~
	\varepsilon_{i,g}\ge\varepsilon_{\min}~\text{(if gateway)},
	~
	\ell_e(t)\le L_{\max}^{\text{class}(e)}
	\Big\}.
\end{align}
This provides a compact and physically grounded representation of connectivity that is used directly by the routing framework in subsequent sections.

\section{Routing Problem Formulation and Proposed Method}
This section establishes a physically consistent yet computationally lightweight routing formulation for optical LEO networks. We emphasize the impact of geometric evolution, jitter-dependent feasibility, and congestion dynamics, and we present a learning-based routing method capable of adapting to fast-changing topology. The formulation is motivated by the fact that optical inter-satellite links (ISLs) are not only distance-limited but also jitter-limited, resulting in a dynamic and constrained connectivity structure \cite{cao2022dynamic}.

\subsection{Dynamic Graph Representation}
At time $t$, the network is modeled as a directed graph
\begin{align}
	\mathcal{G}(t)=\big(\mathcal{V}(t),\mathcal{E}(t)\big),
\end{align}
where $\mathcal{V}(t)$ includes satellites and visible gateways, and $\mathcal{E}(t)$ contains links satisfying geometry and line-of-sight constraints. Because LEO satellites orbit the Earth approximately every 90 minutes, the entire neighbor structure of each satellite changes rapidly. The distance $\ell_e(t)$ on link $e$ is a continuously varying function of time, while elevation masks, plane-crossings, and tracking behavior cause frequent transitions in link availability.

\subsubsection{Intra-Plane vs. Inter-Plane ISLs: A Physics-Based Distinction}
Optical links behave differently depending on the orbital relationship between satellites:
\begin{itemize}
	\item \textbf{Intra-plane ISLs} exhibit slow relative motion and tight pointing stability. Their feasible optical ranges are longer, and their outage probabilities remain low over time.
	\item \textbf{Inter-plane ISLs} experience higher angular rates, leading to more frequent alignment updates and higher jitter variance, thus imposing stricter maximum communication ranges.
\end{itemize}
We incorporate these physical effects using two deterministic thresholds:
\begin{align}
	\ell_e(t)\le
	\begin{cases}
		L_{\max}^{\text{intra}}, & e\in\mathcal{E}_{\text{intra}}, \\[4pt]
		L_{\max}^{\text{inter}}, & e\in\mathcal{E}_{\text{inter}}.
	\end{cases}
	\label{eq:intra_inter_limits}
\end{align}
These values are computed offline from a Gaussian-beam model with pointing jitter and reduce the optical layer to a geometric feasibility filter used directly during routing.

\subsubsection{Busy Nodes and Dynamic Load Constraints}
Satellite relays are subject to finite processing power and buffer capacity. We therefore define a time-varying set of overloaded nodes:
\begin{align}
	\mathcal{V}_{\text{busy}}(t)=\big\{v\in\mathcal{V}(t): Q_v(t)>Q_{\max}\big\},
\end{align}
where $Q_v(t)$ denotes the local queue occupancy. Any link entering a busy node is excluded:
\begin{align}
	e=(i\!\to\!j)\notin\mathcal{E}_{\mathrm{feas}}(t)
	\quad\text{if}\quad j\in\mathcal{V}_{\text{busy}}(t).
\end{align}
This eliminates congestion hot-spots and ensures stable routing even in heavily loaded network states.

\subsection{Latency and Cost Modeling}
Each feasible link $e$ contributes a total cost combining propagation delay and per-hop processing:
\begin{align}
	w_e(t)=\tau_e(t)+\beta
	=\frac{\ell_e(t)}{c}+\beta,
\end{align}
where $c$ is the speed of light and $\beta$ represents typical switching delay. The cumulative cost of a path $\mathcal{P}$ is
\begin{align}
	W(\mathcal{P},t)=\sum_{e\in\mathcal{P}} \left(\frac{\ell_e(t)}{c}+\beta\right).
	\label{eq:full_path_cost}
\end{align}

\subsubsection{Optimization Problem Formulation}
The snapshot routing problem at time $t$ can be formally expressed as:
\begin{align}
	\min_{\mathcal{P}} ~~ & W(\mathcal{P},t)
	\label{eq:opt_problem_full} \\[2pt]
	\text{s.t.} ~~ &
	e\in\mathcal{E}_{\mathrm{feas}}(t), \nonumber \\
	& \ell_e(t) \le L_{\max}^{\text{class}(e)}, \nonumber \\
	& j\notin\mathcal{V}_{\text{busy}}(t) ~~ \forall e=(v\!\to\!j)\in\mathcal{P}, \nonumber \\
	& \mathcal{P}~\text{is a valid directed path between } s_g \text{ and } d_g. \nonumber
\end{align}
This problem captures three interacting constraints:
\begin{enumerate}
	\item \textbf{Dynamic geometry:} all distances $\ell_e(t)$ evolve continuously.
	\item \textbf{Optical feasibility:} jitter-dependent $L_{\max}^{\text{class}(e)}$ restrict ISL availability.
	\item \textbf{Transient congestion:} busy nodes modify the graph topology in real time.
\end{enumerate}
Because $\mathcal{E}_{\mathrm{feas}}(t)$ changes across snapshots and the search is constrained, \eqref{eq:opt_problem_full} is a dynamic constrained shortest-path problem, which is NP-hard in general \cite{kuipers2005conditions}.

\subsection{Learning-Based Routing Framework}
To ensure scalability, we employ a deep reinforcement learning (DRL) policy that makes next-hop decisions using local information only. Unlike conventional shortest-path solvers, a trained DRL agent exploits geometric patterns across many snapshots, enabling real-time adaptation without re-running global graph computations \cite{dong2023drl,chen2025deep,luong2019applications}.

\subsubsection{State Representation}
At each relay $v$, the state vector $s_v(t)$ includes:
\begin{itemize}
	\item \textbf{Geometric distance} to the destination’s serving satellite.
	\item \textbf{Directional improvement:} best neighbor’s reduction in distance.
	\item \textbf{Feasible-degree}: number of neighbors after applying optical and congestion filters.
	\item \textbf{Load-awareness:} local busy indicator or congestion metric.
	\item \textbf{Revisit flag:} preventing loops and stabilizing convergence.
\end{itemize}
This vector captures the essential attributes necessary for decentralized routing under dynamic conditions and matches standard DRL practice for graph-based decision-making \cite{luong2019applications}.

\subsubsection{Action Masking and Feasible Action Set}
The DRL agent may only choose neighbors satisfying all physical and load constraints:
\begin{align}
	\mathcal{A}(v,t)=\Big\{
	(v\!\to\!j):~
	\ell_{v,j}(t) \le L_{\max}^{\text{class}(v,j)},~
	j\notin\mathcal{V}_{\text{busy}}(t)
	\Big\}.
	\label{eq:action_masking}
\end{align}
This masking:
\begin{itemize}
	\item prevents selecting infeasible or outage-prone ISLs,
	\item greatly reduces the decision space,
	\item increases training stability and sample efficiency \cite{luong2019applications}.
\end{itemize}

\subsubsection{State Transition and Episode Termination}
Upon selecting $e=(v\!\to\!j)$, the agent transitions to $j$ and observes a new state $s_j$. Episodes terminate when:
\[
v = d_g \quad \text{or} \quad \mathcal{A}(v,t)=\emptyset.
\]
This snapshot-level model is consistent with real-time LEO routing, where a forwarding decision must be made using the instantaneous connectivity structure \cite{cao2022dynamic}.

\subsection{Complete Routing Algorithm}
Algorithm~\ref{alg:extended_routing} presents the full routing method, including construction of feasible sets, masking, policy selection, and termination conditions.

\begin{algorithm}[t]
	\caption{Snapshot Routing with DRL-Based Feasible-Link Masking}
	\label{alg:extended_routing}
	\begin{algorithmic}[1]
		\State \textbf{Input:} snapshot graph $\mathcal{G}(t)$, source $s_g$, destination $d_g$
		\State \textbf{Output:} routed path $\mathcal{P}$
		\State $\mathcal{P}\gets []$, \quad $v \gets s_g$
		\While{$v \neq d_g$}
		\State compute neighbor set $\mathcal{N}(v,t)$
		\State apply geometric and optical filters using \eqref{eq:intra_inter_limits}
		\State apply congestion filter to remove neighbors in $\mathcal{V}_{\text{busy}}(t)$
		\State form feasible action set $\mathcal{A}(v,t)$ using \eqref{eq:action_masking}
		\If{$|\mathcal{A}(v,t)|=0$}
		\State \textbf{terminate:} \text{no feasible route at snapshot $t$}
		\EndIf
		\State choose next-hop $e=(v\!\to\!j)$ using DRL policy $\pi_\theta$ with masking
		\State append $e$ to $\mathcal{P}$
		\State $v \gets j$
		\EndWhile
		\State \Return $\mathcal{P}$
	\end{algorithmic}
\end{algorithm}

\begin{figure*}
	\centering
	\subfloat[] {\includegraphics[width=2.29 in]{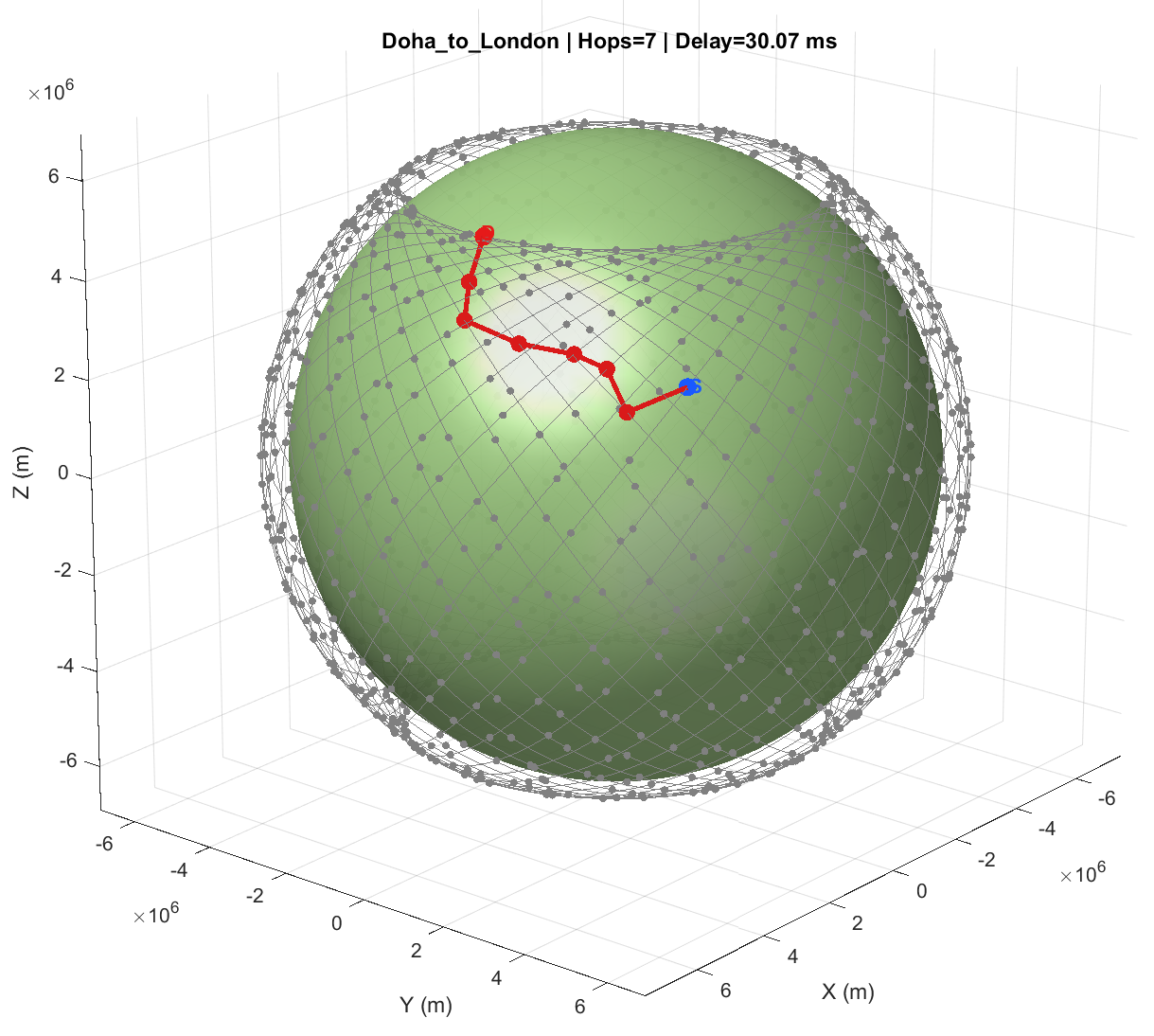}
		\label{cn1}
	}
	\hfill
	\subfloat[] {\includegraphics[width=2.29 in]{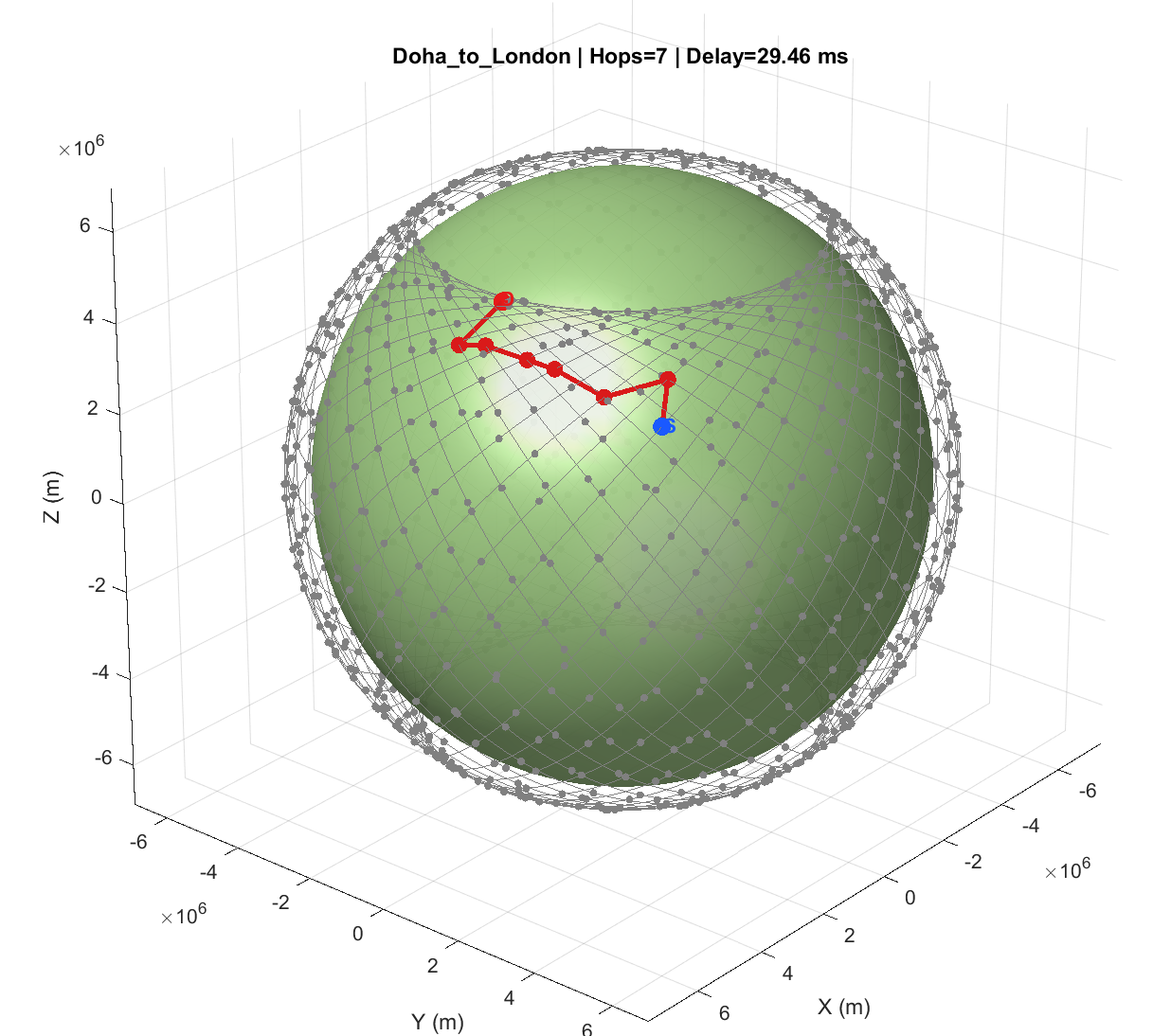}
		\label{cn2}
	}
	\hfill
	\subfloat[] {\includegraphics[width=2.29 in]{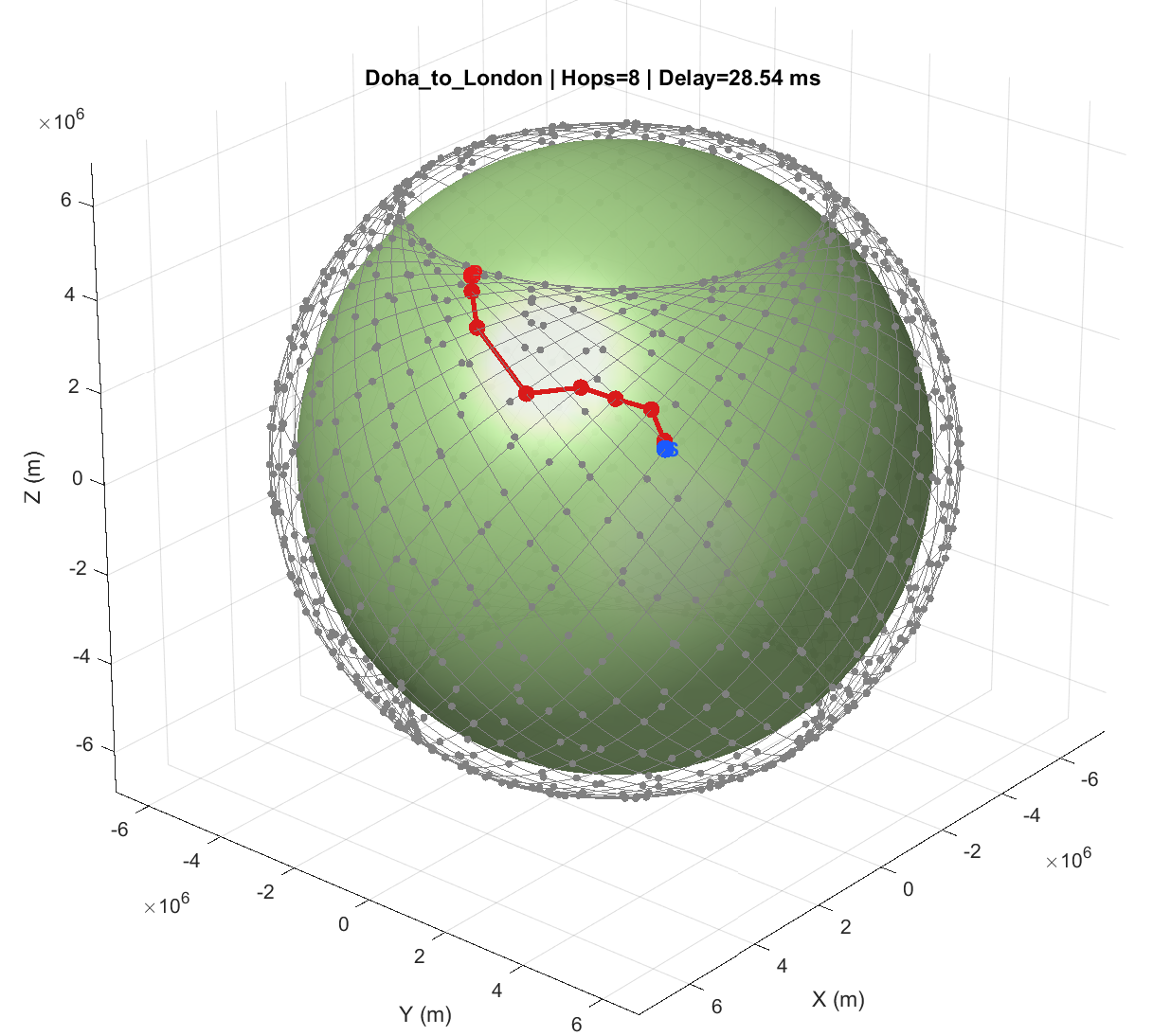}
		\label{cn3}
	}
	\caption{Representative routing paths for the Doha-to-London scenario under different constellation snapshots. Red nodes and links denote the selected route; blue marker indicates the destination satellite.}	
	\label{cb}
\end{figure*}

\section{Simulation Results}
This section presents a concise evaluation of the proposed routing framework on a representative optical LEO constellation. The objective is to illustrate end-to-end routing behavior under geometric constraints, class-dependent ISL feasibility, and pointing-error–limited optical ranges. A simplified Walker-style LEO constellation is generated with uniformly distributed satellites over multiple circular orbital planes. Table~\ref{tab:orbital_params} summarizes the orbital configuration used for all experiments. Each simulation snapshot is obtained by sampling satellite positions at a random orbital time, and ISLs are formed only when the geometric distance satisfies the class-dependent feasibility limits 
\(
\ell_e \le L_{\max}^{\text{intra}}
\)
for intra-plane links and 
\(
\ell_e \le L_{\max}^{\text{inter}}
\)
for inter-plane links.

These range limits are computed analytically from the Gaussian-beam outage model under pointing jitter and represent the maximum distance for which the target outage constraint is satisfied. The divergence angle is not optimized during routing and is absorbed into these feasibility thresholds.

Ground gateways are located at fixed geodetic coordinates, and their serving satellites are determined by minimum Euclidean distance. Only links above the minimum elevation mask are retained.

\begin{table}[t]
	\centering
	\caption{Orbital Parameters Used in Simulation}
	\label{tab:orbital_params}
	\begin{tabular}{l c}
		\hline
		Parameter & Value \\
		\hline
		Orbit altitude $h$ & 550 km \\
		Earth radius $R_e$ & 6371 km \\
		Number of orbital planes & 40 \\
		Satellites per plane & 25 \\
		Inclination & $53^\circ$ \\
		Inter-plane RAAN spacing & Uniform \\
		Snapshot time & Random in $[0,5400]$ s \\
		\hline
	\end{tabular}
\end{table}

Table~\ref{tab:optical_params} lists the optical parameters used to derive the feasible ISL ranges. These values correspond to typical spaceborne coherent optical terminals.

\begin{table}[t]
	\centering
	\caption{Optical ISL Parameters}
	\label{tab:optical_params}
	\begin{tabular}{l c}
		\hline
		Parameter & Value \\
		\hline
		Optical wavelength $\lambda$ & 1550 nm \\
		Transmit power $P_t$ & 1 W \\
		Tx/Rx efficiency $(\eta_t,\eta_r)$ & $(0.5,\,0.5)$ \\
		Receiver aperture radius $a_r$ & 5 cm \\
		System loss & 10 dB \\
		Noise-bandwidth $N_0B$ & $10^{-12}$ W \\
		SNR threshold $\gamma_{\text{th}}$ & 10 (linear) \\
		Max divergence $\theta_{\max}$ & 1 mrad \\
		Pointing jitter $\sigma_\theta$ (intra-plane) & 100 $\mu$rad \\
		Pointing jitter $\sigma_\theta$ (inter-plane) & 200 $\mu$rad \\
		\hline
	\end{tabular}
\end{table}

Using these parameters, the analytically derived range limits for the adopted snapshots are approximately  
\[
L_{\max}^{\text{intra}} \approx 2800~\text{km}, \qquad
L_{\max}^{\text{inter}} \approx 1400~\text{km},
\]
reflecting the higher pointing stability of intra-plane ISLs.

Figs.~\ref{cn1}--\ref{cn3} illustrate three representative routing paths for the Doha-to-London scenario under different constellation snapshots. In each case, the feasible ISL graph is restricted to a narrow angular corridor around the great-circle direction, significantly reducing the action-space complexity while preserving physically reasonable routes.

The learned paths consistently follow a smooth progression along the source--destination geodesic, occasionally exploiting long intra-plane hops when geometrically aligned, and switching to short inter-plane hops when the relative satellite geometry tightens. End-to-end delays are in the range of 28--31~ms with 7--8 hops, demonstrating physically consistent behavior under the distance-constrained ISL model.

Intra-plane ISLs are generally more stable due to small relative angular rates and lower pointing jitter, allowing longer feasible link distances. As a result, the routing policy naturally prefers intra-plane hops when they reduce hop count or provide substantial geometric progress.  
In contrast, inter-plane ISLs have more stringent distance limits due to higher jitter and faster relative motion. Consequently, they appear in the learned routes only when necessary to align with the great-circle path or transition between favorable orbital tracks.

Pointing jitter directly reduces the maximum feasible ISL distance. In our simplified setup, doubling the jitter (e.g., from $100$ to $200~\mu$rad) roughly halves the feasible inter-plane range.  
Across 100 randomly generated snapshots (not shown here), we observed a consistent trend:  
(i) higher jitter increases hop count,  
(ii) end-to-end delay rises modestly, and  
(iii) occasional snapshot failures occur when geometric gaps exceed the jitter-limited range.

\section{Conclusion and Future Work}
This paper introduced a geometry- and QoS-aware routing framework for large optical LEO constellations that integrates class-dependent ISL feasibility, pointing-jitter limitations, and dynamic congestion conditions into a lightweight decision-making model. By representing the constellation through a dynamic feasible graph and enforcing optical constraints via precomputed range limits, routing decisions remain physically consistent without requiring explicit link-budget evaluation during operation. The proposed DRL-based policy, equipped with feasibility-aware action masking, exploits local geometric structure while adapting to rapid changes in link availability. Simulations on a representative Starlink-like orbital configuration demonstrated that the learned routes naturally follow stable intra-plane paths, selectively utilize inter-plane connectors when necessary, and achieve robust end-to-end latency across a wide variety of snapshots.

Several promising extensions can further enhance the proposed framework. First, integrating load-aware link metrics or queueing delay predictions may improve routing performance under heavy traffic. Second, multi-flow or multi-commodity routing formulations could support contention-aware joint optimization for large-scale network operations. Third, extending the physical feasibility model to incorporate terminal switching dynamics, acquisition overhead, or adaptive divergence control would enable finer-grained optical management. Finally, incorporating predictive elements—such as learning-based topology forecasting or anticipatory policy adaptation—may improve performance in scenarios involving highly dynamic gateway selection or long-distance intercontinental routing.

\bibliographystyle{IEEEtran}
\balance
\bibliography{IEEEabrv,myref}

\end{document}